# Change in structural brain network abnormalities after traumatic brain injury determines post-injury recovery

James J Gugger*[,1,2], Nishant Sinha*[,1,2], Yiming Huang[3], Alexa Walter[1], Cillian Lynch[1], Justin Morrison[1], Nathan Smyk[1], Danielle Sandsmark[1], Ramon Diaz-Arrastia†[1], Kathryn A Davis†[1,2]

*Contributed equally as the first author.
†Contributed equally as senior author.

[1]Department of Neurology, Perelman School of Medicine, University of Pennsylvania, Philadelphia, PA, USA

[2]Center for Neuroengineering & Therapeutics, University of Pennsylvania, Philadelphia, PA, USA

[3]Interdisciplinary Computing and Complex BioSystems, School of Computing, Newcastle University, Newcastle upon Tyne, United Kingdom

Correspondence to: Nishant Sinha
Address: Hayden Hall Room 306, University of Pennsylvania, 240 South 33rd Street, Philadelphia, PA 19104, USA
Email: nishant.sinha89@gmail.com
Twitter: @_Nishant_Sinha
Orcid ID: 0000-0002-2090-4889

# Abstract

**Objective:** The trajectory of an individual's recovery after traumatic brain injury (TBI) is heterogeneous, with complete recovery in some cases but persistent disability in others. We hypothesized that changes in structural brain network abnormalities guide the trajectory of an individual's recovery post-injury. Our objective was to characterize the variability in recovery post-TBI by identifying a putative neuroimaging biomarker of traumatic axonal injury (TAI) in individuals with mild TBI.

**Method:** We analyzed 70 T1-weighted and diffusion-weighted MRIs longitudinally collected from 35 individuals during the subacute and chronic post-injury periods. Each individual underwent longitudinal blood work to characterize blood protein biomarkers of axonal and glial injury and assessment of post-injury recovery in the subacute and chronic periods. By comparing the MRI data of individual cases with 35 controls, we estimated the longitudinal change in structural brain network abnormalities. We validated this proxy measure of TAI with independent measures of acute intracranial injury estimated from head CT and blood protein biomarkers. Using elastic net regression models, we identified brain regions in which the change in abnormality over time correlated with the change in symptom burden.

**Results:** Post-injury structural network abnormality was significantly higher than controls in both subacute and chronic periods, associated with an acute CT lesion and subacute blood levels of glial fibrillary acid protein (r=0.5, p=0.008) and neurofilament light (r=0.41, p=0.02). Longitudinal change in abnormality associated with change in functional outcome status (r=-0.51, p=0.003) and post-concussive symptoms (BSI: r=0.46, p=0.03; RPQ:r = 0.46, p=0.02). Brain regions that most closely mapped onto symptom change over time corresponded to structural network hubs or areas susceptible to neurotrauma.

**Conclusion:** Structural network abnormalities might be a biomarker of TAI both at the resolution of the whole brain and in individual brain regions. Assessing changes in brain network abnormality over time might enable better patient stratification for monitoring recovery after neurotrauma.

# Introduction

Traumatic brain injury (TBI) is a global public health issue, and even its most mild form—mild TBI (mTBI) or concussion—carries a significant risk of chronic neurological complications such as dementia[1] and epilepsy[2,3]. TBI results in injury to neurons, glia, and cerebral blood vessels with substantial variability in severity and anatomic location between individuals. This mechanistic and spatial heterogeneity likely plays a large role in the substantial inter-individual variability in outcomes. Although acute CT and MRI findings such as the presence of traumatic subarachnoid hemorrhage and contusions influence outcome after TBI,[4] efforts at developing targeted therapies for preventing or mitigating the most severe complications of TBI, such as dementia and epilepsy[5] requires more sophisticated methods of risk stratification.

The most common and well-described pathologic feature of TBI is traumatic axonal injury (TAI).[6,7] TAI is responsible for much of the long-term disability after neurotrauma[7] and may incite the process of progressive neurodegeneration[8]. Disability resulting from TAI may be a consequence of injury to critical axonal tracts, such as brainstem reticular pathways,[9] but more commonly is associated with the aggregate burden of disconnection in widely distributed neural networks.[10,11] Indeed, some white matter tracts are more vulnerable than others due to their orientation and location, such as the midline structures (corpus callosum, fornix, and cingulum) or the gray-white junction between tissue compartments[11,12,13,14]. While TAI was initially described in the setting of severe TBI and prolonged coma, it is now widely recognized that TAI is a predominant mechanism of disability in mild and moderate TBI[10]. Thus, TAI could be the mechanism underlying post-traumatic neurodegeneration and epileptogenesis. Unfortunately, there are currently no biomarkers that can characterize the individual burden of TAI.

TAI can be characterized *in vivo* by diffusion MRI, with most studies reporting a disruption of white matter microstructural integrity in TBI.[15] Numerous methods exist for the analysis of diffusion MRI data; however, current analytic methods fail to adequately account for both aggregate and region-specific abnormalities in structural connectivity. Normative modeling is a promising case-control approach that quantifies abnormalities in individual cases as the deviations from a matched cohort of healthy individuals. Recent application of normative modeling characterized both the whole brain as well as region-specific structural network abnormalities that predicted surgical outcomes in focal epilepsy[16,17]. This approach is a promising tool for characterizing the overall burden and spatial pattern of TAI but has not yet been applied in TBI. An aggregate measure of TAI would be useful as a biomarker in future therapeutic trials of agents targeting TAI. Tracking changes in the aggregate burden of TAI may be useful as a prognostic biomarker by identifying individuals who may not fully recover after TBI.

The main objective of this study was to assess the association between longitudinal change in traumatic axonal injury with the trajectory of functional recovery and post-injury symptom severity. We hypothesized that improvement in traumatic axonal injury measured between the subacute and chronic phases would be associated with an increased likelihood of clinical recovery after TBI. To test this hypothesis, we quantified traumatic axonal injury as abnormalities on structural brain imaging data acquired during the subacute and chronic phases after TBI. We reconciled injury heterogeneity in individual cases at the whole-brain level as the net burden/load of abnormalities by applying a normative modeling approach. We validated this proxy measure of traumatic axonal injury against two complementary modalities used clinically to assess injury severity: a) abnormalities on CT images and b) expression of blood protein biomarkers of axonal and glial injury. We found the change in abnormality load associated with the change in functional outcome and burden of post-injury symptoms after TBI. As potential therapeutic targets, our study identified the most vulnerable brain regions in which the change in abnormality over time guided the trajectory of individual recovery post-injury.

## Materials and Methods

### Participants

Thirty-five adult individuals with TBI who were hospitalized as a result of their injury were enrolled at a Level 1 Trauma Center. Individuals were excluded for a history of pre-existing serious neurological or psychiatric disorder, comorbid disabling condition limiting outcome assessment, current pregnancy, or if they were incarcerated. Thirty-five demographically matched healthy control subjects recruited from the general population with no history of TBI, pre-existing disabling neurological or psychiatric disorder, or current pregnancy were also enrolled. Individuals with TBI underwent a clinical CT imaging upon admission and a research protocol including brain MRI in the subacute and chronic post-injury period and longitudinal blood draws at multiple time points, including subacute and chronic post-injury periods. The subacute period was approximated at 14 days (2 weeks), and the chronic period was approximated at 180 days (6 months) after injury. Healthy controls underwent a single brain MRI. Demographic information, medical history, admission injury characteristics, and other clinical information were collected from the medical records. These details and timelines are illustrated in Figure 1a. This study was approved by the Institutional Review Board at the University of Pennsylvania, and written consent was obtained by the individual or a legally authorized representative when appropriate.

[Figure 1]

### Acute Head CT Interpretation

Each participant's CT scan and the accompanying clinical neuroradiologist's report were independently evaluated by a board-certified neurologist (JG) for the presence or absence of acute intracranial lesions. An acute intracranial lesion was defined as any trauma-related finding, including epidural hematomas, subdural hematomas, indeterminate extra-axial lesions, contusions, intraventricular hemorrhage, subarachnoid hemorrhage, and petechial hemorrhage.[18] Presence of any of the above intracranial lesions was considered CT-positive; otherwise, the individual was categorized as CT-negative. Consistent with prior studies,[18] isolated skull fracture was not considered an acute intracranial lesion.

**Measurement of Neurodegeneration-Associated Proteins in Blood Serum**

Neurofilament light (NfL) is a protein component of the neuronal cytoskeleton and a putative biomarker of axonal injury that is released into the CSF and blood in several neurodegenerative diseases, including TBI.[19] Glial fibrillary acidic protein (GFAP) is a protein expressed almost exclusively in astrocytes and is released into the CSF and blood following the disintegration of the astrocyte cytoskeleton.[20] Both proteins can be measured in the blood, and blood concentrations are associated with both neuroimaging abnormalities[21,22] and outcomes after TBI.[23,24] To establish the relationship between structural network abnormalities and blood markers of TAI, we collected blood samples via venipuncture within 24 hours, 72 hours, two weeks, three months, and six months of injury. NfL and GFAP were measured on the Quanterix HD-X Analyzer™ (Quanterix, Billerica, MA) via Single Molecule Assay (Simoa®), using the 'Neurology 4-plex B' immunoassay kits.

**Outcome assessment**

Post-injury recovery was assessed at the same time as the MRI using three measures: self-reported symptomatology was assessed with the Rivermead Post-Concussion Symptom Questionnaire (RPQ)[25] and the Brief Symptom Inventory-18 (BSI-18), while the post-injury functional outcome was assessed using the Glasgow Outcome Scale-Extended (GOSE).[26] Outcome assessment was conducted by trained research personnel who were blind to imaging and biomarker findings. To understand the relationship between the temporal profile of neuroimaging abnormalities and post-TBI symptom burden, we derived a measure that includes information at both time points as well as their change over time. This is defined as proportional change. For measures of symptom severity, the proportional change is defined as the following:

$$change\ in\ outcome\ score = \frac{score(chronic) - score(subacute)}{score(chronic) + score(subacute)}$$

**Neuroimaging data acquisition and preprocessing**

Brain MRIs were performed on a 3T scanner (Siemens Prisma) using a product 32-channel head coil. Structural imaging included a sagittal T1-weighted MPRAGE (TR = 2.3 s, TE = 2.94 ms, TI

= 900 ms, FA = 9°, resolution = 1 x 1 x 1 mm). Whole brain diffusion MRI was performed with an echo planar sequence with FA 90° and resolution = 2.4 x 2.4 x 2.4 mm (b-value = 1000 s/mm$^2$, 64 diffusion directions, TR 2.9 s, TE 94 ms). Separately, 14 images with b-value = 0 s/mm$^2$ with reverse phase encoding were acquired. Diffusion MRI data were corrected for eddy current and movement artifacts using the FSL topup and eddy tool.[27] The diffusion data for each subject was registered and reconstructed to the standard ICBM-152 space using q-space diffeomorphic reconstruction implemented in DSI studio (http://dsi-studio.labsolver.org).

**Measurement of Structural Network Abnormalities**

The process for delineating structural network abnormalities is outlined in Figure 1. For each participant, we constructed a structural brain network consisting of gray matter regions/nodes and white matter connections (edges) between regions. We defined 90 contiguous cortical and subcortical regions from the AAL atlas. To identify the connectivity between regions, we performed whole-brain tracking using deterministic tractography. To avoid any bias, we applied identical fiber tracking parameters on each scan across all subjects. Tractography generated approximately 1,000,000 tracts with tracking parameters configured as follows: Runge-Kutta method with step size 1mm, whole-brain seeding, initial propagation direction set to all fiber orientations, minimum tract length 15mm, maximum tract length 300mm, and topology informed pruning applied with one iteration to remove false connections. The whole-brain structural network is constructed by delineating connections between all pairs of gray matter regions of interest (ROIs) in the parcellation scheme. Using this schema, a structural connection exists between two regions if they are linked by a streamline. We weighted the connectivity across all streamlines that connect each pair of regions by tract counts normalized by the average volume of the regions connected by those streamlines.[28] Repeating this process for each pair of nodes resulted in weighted connectivity matrices of size 90 x 90 per participant, where the x- and y-axes of the matrix are the regions identified by the brain parcellation scheme, and the elements of the matrix are populated with the white matter streamlines connecting each pair of regions (Figure 1b-d).

After structural network construction, we computed a z-score for each connection between regions in each TBI subject using the mean and standard deviation of the respective connection in the uninjured controls. A connection was considered abnormal if the z-score exceeded a specific threshold. To identify the appropriate z-score threshold above which a node is considered abnormal, we varied the z-score threshold from 2 to 3.5 in increments of 0.1. Node abnormality is the degree of this thresholded network, and abnormality load is the total number of abnormal nodes in the network. These steps are illustrated in Figure 1e-g. In the main manuscript, we show the result at one example threshold at $z > 2.5$, whereas in supplementary Figures S1 and S2, we show consistent results across all thresholds and alternative atlas of brain parcellation.

Similar to the proportional change for symptom severity, we calculated the proportional change of post-TBI abnormality load:

$$change\ in\ abnormality = \frac{abnormality\ load(chronic) - abnormality\ load(subacute)}{abnormality\ load(chronic) + abnormality\ load(subacute)}$$

**Elastic net regression model**

To identify structural network nodes important for recovery, we mapped the relationship between change in abnormality among the 90 individual nodes over time with recovery measures, incorporating feature selection with elastic net regression.[29] Elastic net regression is a regularization technique that permits the identification of important predictors in the presence of numerous redundant or highly correlated predictors. Elastic net is a combination of lasso (L1 norm) and ridge (L2 norm) regression aiming to optimize the following objective function:

$$min_{\beta_0,\beta} \left( \frac{1}{2N} \sum_{i=1}^{N}(y_i - \beta_o - x_i^T\beta)^2 + \lambda P_\alpha(\beta) \right),$$

where,

$$P_\alpha(\beta) = \frac{(1-\alpha)}{2}\|\beta\|_2^2 + \alpha\|\beta\|_1$$

In this objective function, $N$ is the number of observations, $y_i$ is the response at observation $i$, $x_i$ is the input data or feature vector of length $p$ at observation $i$, and parameters $\beta_o$, $\beta$ are the scalar intercept and weight vector of length $p$ which are estimated by the model training process. $\lambda$ (lambda) and $\alpha$ (alpha) are the regularization parameters that scale the L1 ($\|\beta\|_1$) and L2 ($\|\beta\|_2^2$) norms to prevent overfitting of the regression model.

We applied principal component analysis to the proportional change of the three measures of recovery and extracted the first principal component. The first principal component accounted for 66% of the variance of the three measures of recovery, which we incorporated as the response variable ($y_i$) in the model. The proportional change in node abnormality for each of the 90 network nodes was entered as a feature vector ($x_i$ of length $p$ = 90). During model training, parameter $\beta$ weighted each network node, which corresponded directly to the relative importance of the change in abnormality of that node in predicting the change in recovery status of a patient. The regularization parameters $\lambda$ (lambda) and $\alpha$ (alpha) guided the feature selection process by identifying a minimum number of nodes from 90 nodes (features) that increases the prediction performance of the regression model on held-out data. We incorporated a nested cross-validation scheme for model training, tuning, and testing. In the outer fold, we implemented a leave-one-out cross-validation scheme to test the model prediction accuracy. In the inner fold, we incorporated a three-fold cross-validation scheme for model training and tuning. We trained the model on the training fold and tuned the model with different regularization parameters on the inner cross-validation folds. To select the appropriate regularization parameters, we set α=0.5, and performed a grid search varying $\lambda$ over values ranging from 0.001 to 1.[16,30] We selected the model with the lowest mean square error for the testing model prediction performance on unseen test data from the outer leave-one-out cross-validation fold. The normalized coefficients of the resultant model indicated the relative importance of the individual node for predicting patient recovery.

**Statistical Analysis**

To investigate whether an increase in abnormality load is associated with traumatic brain injury, we applied a non-parametric Wilcoxon rank-sum test. A one-tailed *p* value was computed with the ranksum function in MATLAB (MathWorks, Inc, Natick, MA), incorporating the exact method. Results are declared significant for *p<0.05*. We applied Benjamini-Hochberg false discovery rate correction at a significance level of 5%. To assess change in abnormality load in TBI cases from subacute to chronic period, we applied paired statistics, testing significance using signed ranksum test. Effect size between groups was computed with the Cohen's d score, and the correlation coefficients between blood biomarkers and abnormality load as well as between proportional change in abnormality load and measures of clinical outcomes were determined with the Spearman rank order. We computed 95% bootstrap confidence intervals (CIs) of effect size using a bias-corrected and accelerated percentile method from 10,000 bootstraps resamples with replacement. We did not impute missing values to avoid introducing any bias. Chi-squared or Fisher's exact test was used for comparing categorical data between groups.

**Data availability**

De-identified clinical and imaging data will be made available upon request to the corresponding author.

# Results

## Clinical Data

The cohort included 35 TBI participants with a median admission Glasgow Coma Scale (GCS) of 15 (interquartile range (IQR) 14-15) scanned in the subacute (median 16 days after TBI, IQR 13-25 days) and chronic (median 195 days, IQR 188-214) post-injury period, as well as age and sex, matched 35 healthy controls scanned at a single time point. Demographic information and injury characteristics for all participants are shown in Table 1. As expected, a significant change at the group level in GFAP, NfL, BSI, RPQ, and GOSE scores between the subacute and chronic period indicates the trajectory towards recovery and the validity of these markers and instruments in assessing outcomes in individuals after TBI. We detected no significant association between change in blood plasma concentration of GFAP and NfL with change in outcome scores (BSI, RPQ, and GOSE) from subacute to chronic periods (Figure S3).

[Table 1]

## Abnormality load, acute CT findings, and blood biomarkers

Individuals with TBI had significantly higher abnormality load compared to controls in both the subacute ($p = 0.03, d = 0.51$) and chronic period ($p = 0.04, d = 0.48$) (Figure 2a). There was no difference between abnormality load in the subacute and chronic post-injury period. Abnormality load in the subacute period was significantly higher among TBI cases, with evidence of an acute intracranial lesion on initial head CT (Figure 2b). Subacute abnormality load correlated with serum GFAP ($r = 0.5, p = 0.008$) and NFL ($r = 0.41, p = 0.02$) measured in the subacute period (Figure 2c,d). Chronic abnormality load did not correlate with serum GFAP or NfL measured in the chronic period. Figure S4 shows that on average more links have an abnormally high z-score for the TBI group both in subacute and chronic periods than the control group.

[Figure 2]

## Abnormality load correlations with outcomes

The proportional change in abnormality load from the subacute to chronic post-injury period was 0.002 (95% CI -0.1, 0.01), indicating that at the group level, abnormality load did not significantly change over time; however, abnormality load increased over time for 19 patients while it decreased for 15 patients and remain unchanged in one patient (Figure 2a). Figure 3 shows the correlation between proportional change in abnormality load and proportional change in measures of recovery. The proportional change in abnormality load correlated with measures of subjective reporting and functional outcome. There was a positive correlation between proportional change in abnormality load and the BSI and RPQ ($r = 0.46$, $p = 0.03$ for BSI; $r = 0.46$, $p = 0.02$ for RPQ). This indicates that increases in structural network abnormality over time correlate with worsening post-concussive symptoms over time. There was a negative correlation between proportional change in abnormality load and functional outcome as assessed by the Glasgow Outcome Scale-Extended (GOSE) ($r = -0.51$, $p = 0.003$). This indicates that increases in structural network abnormality over time correlates with worsening functional outcome over time.

[Figure 3]

**Node abnormality and outcome**

To delineate the relationship between proportional change in each node over time with a proportional change in recovery status, we utilized elastic net regression. First, we performed a principal component analysis on the proportional change in the BSI, RPQ, and GOSE and extracted the first principal component score, which explained 66% of the variance in the three measures. The model included the first principal component score as the response variable (a measure of global recovery) and 90 features corresponding to the proportional change in abnormality of each node in the structural brain network. We incorporate a leave-one-out cross-validation scheme for model testing and three-fold cross validation for model training and tuning. For each choice of the regularization parameter, the mean square error from the model training and tuning process is shown in Figure 4a. The results of the model with the lowest mean square error are shown in Figure 4b-d. Figure 4b depicts the normalized coefficients of the 14 nodes in the model for every choice of lambda, and the red line depicts the model with the lowest mean square error. The correlation between the predicted and actual recovery score was -0.75 (95% CI -0.5, -0.87), p < 0.005, Figure 4c. Figure 4d depicts the location of the 14 structural network nodes included in the final model; they corresponded to regions previously identified as structural network hubs (i.e., visual cortex, basal ganglia, cingulate cortex) or areas susceptible to TBI (i.e., inferior frontal cortex).[31,32] Supplementary Table S1 shows the node names alongside their normalized regression coefficient.

[Figure 4]

# Discussion

The main findings in this study are that: (1) whole brain structural network abnormality load is greater in individuals with TBI compared to healthy controls, (2) in individual TBI cases, the burden of abnormality can increase, decrease, or remain stable from the subacute to chronic post-injury period, (3) abnormality load measured in the subacute post-injury period is significantly higher in those with acute intracranial lesions visible on CT, (4) abnormality load correlates with serum concentrations of GFAP and NfL in the subacute period, (5) the proportional change in abnormality load correlates with the change in the severity of self-reported symptoms and functional outcome, and (6) change in node abnormality of structural network hubs is related to recovery status after TBI. Collectively, these findings support the notion that structural network abnormalities show promise as a measure of traumatic axonal injury (TAI).

Abnormality load is a summary score describing deviations in whole brain structural connectivity. Understanding the aggregate level of deviation from normal structural connectivity can provide an estimation of the degree of TAI. This has important implications for long-term neurocognitive outcomes, given the association between the extent of TAI and the progression of neurodegeneration after moderate-to-severe TBI.[8] Numerous studies have examined longitudinal changes in white matter microstructural integrity after TBI using diffusion MRI.[11,15,33–38] In general, while there is some variability, most studies consistently show greater degrees of abnormalities in those with greater injury severity. On the contrary, few have specifically examined longitudinal changes in whole brain structural connectivity. Silva et al.[35] and Dall'Acqua et al.[39] analyzed longitudinal changes in graph theory measures of structural connectivity in patients with primarily mild TBI and found that dynamic changes in connectivity over time are associated with patterns of cognitive recovery; however, they did not report how to aggregate deviations in structural connectivity changed over time. In a cross-sectional analysis, Taylor et al.[40] reported a novel measure of injury severity in a cohort of patients with mild TBI by selecting 22 white matter tracts inferred from diffusion tractography and calculating the Mahalanobis distance, which represents an aggregate score describing abnormal connectivity. They found that the Mahalanobis distance was able to differentiate patients with mild TBI from healthy controls and that it is correlated with overall cognitive function. Our study represents the first study to report on deviations in whole brain structural connectivity in TBI.

We found that whole brain abnormality load in the subacute post-injury period after mild TBI correlates with concentrations of NfL and GFAP. NfL and GFAP, measured in the acute and subacute period after TBI, predict neurologic outcomes at six months and one year.[23,24] NfL is thought to be a marker of axonal injury and is elevated across the spectrum of neurodegenerative diseases.[19] GFAP is thought to be a measure of glial injury; however, one study showed that GFAP was able to distinguish patients with MRI signs of diffuse axonal injury compared with patients with other lesion types.[41] Our results are in agreement with prior work, as prior studies demonstrated more robust correlations between NfL and GFAP concentrations with microstructural integrity of the corpus collosum than tau and UCHL1.[21] For all blood biomarkers examined, we observed peak concentrations during the acute-to-subacute period after TBI followed by declining concentrations in the chronic period. Conversely, abnormality load did not appreciably change between the subacute and chronic periods. This could reflect acute traumatic axonal injury with the acute release of proteins into the bloodstream followed by declining concentrations in the blood over time, while injured white matter tracts remain disrupted over time as reflected by stability in imaging. How the process of axonal injury and repair is reflected in blood and imaging biomarker profiles during this dynamic process remains to be determined.

We found that change in node abnormality of structural network hubs is related to recovery status after TBI. These findings are in keeping with prior work demonstrating a relationship between structural network hubs and cognition in patients with TBI. Similar to our study, Fagerholm et al.[30] used elastic net regression to demonstrate that betweenness centrality and eigenvector centrality, two graph theoretical measures of hub status, are significantly associated with scores on tests of information processing speed, executive function, and associative memory. Our findings and the results of Fagerholm et al. suggest that recovery after TBI is related to the effects of TAI on structural network hubs.

The findings of this work should be interpreted in the context of several limitations. First, patients were scanned at two different, non-uniform time points from the subacute and chronic post-injury period. Structural connectivity changes after TBI likely reflect a dynamic and interactive process of injury, and repair and measurement at two varying time points are unlikely to encapsulate these changes. Second, although at a group level, abnormality load did not change over time, it is clear that there is considerable variability where some patients structural network abnormalities progressed, whereas others became more similar to healthy controls. We did not have the statistical power to analyze the factors underlying the progressive improvement or deterioration in structural networks. Future studies using longitudinal neuroimaging data from large multicenter studies will be necessary to delineate the factors underlying the diverse trajectory of structural network abnormalities.

We have shown significant multifocal brain structural network abnormalities in patients with TBI from the subacute to chronic post-injury period. These abnormalities correlate with blood biomarkers of axonal injury as well as recovery after TBI. These findings indicate that structural network abnormality measurements represent a promising method for the assessment of TAI both at the resolution of the whole brain as well as in individual brain regions. These findings require confirmation in larger longitudinal studies. Nonetheless, this study represents an important step toward developing a non-invasive biomarker that could be used in clinical trials to identify the burden of TAI, monitor the process of post-traumatic epileptogenesis and neurodegeneration, and better patient stratification to track recovery after neurotrauma.

# Acknowledgments

This work was supported by the Pennsylvania Department of Health and was supported by NINDS U01NS086090, U01NS114140, W81XWH-12-2-0139, W81XWH-14-2-0176, and W81XWH-19-2-0002. JG acknowledges support from NINDS T32NS091006 and the American Epilepsy Society/Citizens United for Research in Epilepsy (Research and Training Fellowship for Clinicians). NS acknowledges support from NINDS R01NS116504. KAD acknowledges support from NINDS R01NS116504, R01NS110347, R56NS099348.

# References

1. Barnes DE, Byers AL, Gardner RC, Seal KH, Boscardin WJ, Yaffe K. Association of mild traumatic brain injury with and without loss of consciousness with dementia in US military veterans. *JAMA Neurol*. 2018;75(9):1055-1061. doi:10.1001/jamaneurol.2018.0815
2. Pugh MJ V., Orman JA, Jaramillo CA, et al. The prevalence of epilepsy and association with traumatic brain injury in veterans of the Afghanistan and Iraq wars. *J Head Trauma Rehabil*. 2015;30(1):29-37. doi:10.1097/HTR.0000000000000045
3. Pugh MJ, Kennedy E, Gugger JJ, et al. The Military Injuries: Understanding Post-Traumatic Epilepsy Study: Understanding Relationships among Lifetime Traumatic Brain Injury History, Epilepsy, and Quality of Life. *J Neurotrauma*. 2021;38(20):2841-2850. doi:10.1089/neu.2021.0015
4. Yuh EL, Mukherjee P, Lingsma HF, et al. Magnetic resonance imaging improves 3-month outcome prediction in mild traumatic brain injury. *Ann Neurol*. 2013;73(2):224-235. doi:10.1002/ana.23783
5. Diaz-Arrastia R, Kochanek PM, Bergold P, et al. Pharmacotherapy of traumatic brain injury: State of the science and the road forward: Report of the department of defense neurotrauma pharmacology workgroup. *J Neurotrauma*. 2014;31(2):135-158. doi:10.1089/neu.2013.3019
6. Smith DH, Meaney DF, Shull WH. Diffuse Axonal Injury in Head Trauma. *J Head Trauma Rehabil*. 2003;18(4):307-316. doi:10.1097/00001199-200307000-00003
7. Johnson VE, Stewart W, Smith DH. Axonal pathology in traumatic brain injury. *Exp Neurol*. 2013;246:35-43. doi:10.1016/j.expneurol.2012.01.013
8. Graham NSN, Jolly A, Zimmerman K, et al. Diffuse axonal injury predicts neurodegeneration after moderate-severe traumatic brain injury. *Brain*. 2020;143(12):3685-3698. doi:10.1093/brain/awaa316
9. Edlow BL, Haynes RL, Takahashi E, et al. Disconnection of the Ascending Arousal System in Traumatic Coma. *J Neuropathol Exp Neurol*. 2013;72(6):505-523. doi:10.1097/NEN.0b013e3182945bf6
10. Johnson VE, Stewart JE, Begbie FD, Trojanowski JQ, Smith DH, Stewart W. Inflammation and white matter degeneration persist for years after a single traumatic brain

injury. *Brain*. 2013;136(1):28-42. doi:10.1093/brain/aws322

11. Wang JY, Bakhadirov K, Abdi H, et al. Longitudinal changes of structural connectivity in traumatic axonal injury. *Neurology*. 2011;77(9):818-826. doi:10.1212/WNL.0b013e31822c61d7

12. Wang JY, Bakhadirov K, Devous MD, et al. Diffusion Tensor Tractography of Traumatic Diffuse Axonal Injury. *Arch Neurol*. 2008;65(5). doi:10.1001/archneur.65.5.619

13. Cloots RJH, van Dommelen JAW, Kleiven S, Geers MGD. Multi-scale mechanics of traumatic brain injury: predicting axonal strains from head loads. *Biomech Model Mechanobiol*. 2013;12(1):137-150. doi:10.1007/s10237-012-0387-6

14. Cloots RJH, Gervaise HMT, van Dommelen JAW, Geers MGD. Biomechanics of Traumatic Brain Injury: Influences of the Morphologic Heterogeneities of the Cerebral Cortex. *Ann Biomed Eng*. 2008;36(7):1203-1215. doi:10.1007/s10439-008-9510-3

15. Wallace EJ, Mathias JL, Ward L. Diffusion tensor imaging changes following mild, moderate and severe adult traumatic brain injury: a meta-analysis. *Brain Imaging Behav*. 2018;12(6):1607-1621. doi:10.1007/s11682-018-9823-2

16. Sinha N, Wang Y, Moreira da Silva N, et al. Structural Brain Network Abnormalities and the Probability of Seizure Recurrence After Epilepsy Surgery. *Neurology*. 2021;96(5):e758-e771. doi:10.1212/WNL.0000000000011315

17. Sinha N, Peternell N, Schroeder GM, et al. Focal to bilateral tonic–clonic seizures are associated with widespread network abnormality in temporal lobe epilepsy. *Epilepsia*. 2021;62(3):729-741. doi:10.1111/epi.16819

18. Yuh EL, Jain S, Sun X, et al. Pathological Computed Tomography Features Associated with Adverse Outcomes after Mild Traumatic Brain Injury: A TRACK-TBI Study with External Validation in CENTER-TBI. *JAMA Neurol*. 2021;78(9):1137-1148. doi:10.1001/jamaneurol.2021.2120

19. Bacioglu M, Maia LF, Preische O, et al. Neurofilament Light Chain in Blood and CSF as Marker of Disease Progression in Mouse Models and in Neurodegenerative Diseases. *Neuron*. 2016;91(1):56-66. doi:10.1016/j.neuron.2016.05.018

20. Diaz-Arrastia R, Wang KKW, Papa L, et al. Acute biomarkers of traumatic brain injury: Relationship between plasma levels of ubiquitin C-terminal hydrolase-l1 and glial fibrillary acidic protein. *J Neurotrauma*. 2014;31(1):19-25. doi:10.1089/neu.2013.3040

21. Shahim P, Politis A, van der Merwe A, et al. Time course and diagnostic utility of NfL, tau, GFAP, and UCH-L1 in subacute and chronic TBI. *Neurology*. 2020;95(6):e623-e636. doi:10.1212/WNL.0000000000009985
22. Shahim P, Politis A, van der Merwe A, et al. Neurofilament light as a biomarker in traumatic brain injury. *Neurology*. 2020;95(6):e610-e622. doi:10.1212/WNL.0000000000009983
23. Lei J, Gao G, Feng J, et al. Glial fibrillary acidic protein as a biomarker in severe traumatic brain injury patients: A prospective cohort study. *Crit Care*. 2015;19(1):362. doi:10.1186/s13054-015-1081-8
24. Vos PE, Jacobs B, Andriessen TMJC, et al. GFAP and S100B are biomarkers of traumatic brain injury: An observational cohort study. *Neurology*. 2010;75(20):1786-1793. doi:10.1212/WNL.0b013e3181fd62d2
25. King NS, Crawford S, Wenden FJ, Moss NEG, Wade DT. The Rivermead Post Concussion Symptoms Questionnaire: a measure of symptoms commonly experienced after head injury and its reliability. *J Neurol*. 1995;242(9):587-592. doi:10.1007/BF00868811
26. Jennett B, Snoek J, Bond MR, Brooks N. Disability after severe head injury: observations on the use of the Glasgow Outcome Scale. *J Neurol Neurosurg Psychiatry*. 1981;44(4):285-293. doi:10.1136/jnnp.44.4.285
27. Jenkinson M, Beckmann CF, Behrens TEJ, Woolrich MW, Smith SM. FSL. *Neuroimage*. 2012;62(2):782-790. doi:10.1016/j.neuroimage.2011.09.015
28. Taylor PN, Han CE, Schoene-Bake J-C, Weber B, Kaiser M. Structural connectivity changes in temporal lobe epilepsy: Spatial features contribute more than topological measures. *NeuroImage Clin*. 2015;8:322-328. doi:10.1016/j.nicl.2015.02.004
29. Zou H, Hastie T. Regularization and variable selection via the elastic net. *J R Stat Soc Ser B (Statistical Methodol*. 2005;67(2):301-320. doi:10.1111/j.1467-9868.2005.00503.x
30. Fagerholm ED, Hellyer PJ, Scott G, Leech R, Sharp DJ. Disconnection of network hubs and cognitive impairment after traumatic brain injury. *Brain*. 2015;138(6):1696-1709. doi:10.1093/brain/awv075
31. Oldham S, Fornito A. The development of brain network hubs. *Dev Cogn Neurosci*. 2019;36(June 2018):100607. doi:10.1016/j.dcn.2018.12.005

32. Crossley NA, Mechelli A, Scott J, et al. The hubs of the human connectome are generally implicated in the anatomy of brain disorders. *Brain*. 2014;137(8):2382-2395. doi:10.1093/brain/awu132
33. Palacios EM, Owen JP, Yuh EL, et al. The evolution of white matter microstructural changes after mild traumatic brain injury: A longitudinal DTI and NODDI study. *Sci Adv*. 2020;6(32):eaaz6892. doi:10.1126/sciadv.aaz6892
34. Johnson EL, Krauss GL, Lee AK, et al. Association between white matter hyperintensities, cortical volumes, and late-onset epilepsy. *Neurology*. 2019;92(9):E988-E995. doi:10.1212/WNL.0000000000007010
35. Moreira da Silva N, Cowie CJA, Blamire AM, Forsyth R, Taylor PN. Investigating Brain Network Changes and Their Association With Cognitive Recovery After Traumatic Brain Injury: A Longitudinal Analysis. *Front Neurol*. 2020;11(June):1-11. doi:10.3389/fneur.2020.00369
36. Wang JY, Bakhadirov K, Devous MD, et al. Diffusion tensor tractography of traumatic diffuse axonal injury. *Arch Neurol*. 2008;65(5):619-626. doi:10.1001/archneur.65.5.619
37. Warner MA, De La Plata CM, Spence J, et al. Assessing spatial relationships between axonal integrity, regional brain volumes, and neuropsychological outcomes after traumatic axonal injury. *J Neurotrauma*. 2010;27(12):2121-2130. doi:10.1089/neu.2010.1429
38. Marquez de la Plata CD, Yang FG, Wang JY, et al. Diffusion Tensor Imaging Biomarkers for Traumatic Axonal Injury: Analysis of Three Analytic Methods. *J Int Neuropsychol Soc*. 2011;17(01):24-35. doi:10.1017/S1355617710001189
39. Dall'Acqua P, Johannes S, Mica L, et al. Functional and structural network recovery after mild traumatic brain injury: A 1-year longitudinal study. *Front Hum Neurosci*. 2017;11(May):1-16. doi:10.3389/fnhum.2017.00280
40. Taylor PN, Moreira Da Silva N, Blamire A, Wang Y, Forsyth R. Early deviation from normal structural connectivity: A novel intrinsic severity score for mild TBI. *Neurology*. 2020;94(10):e1021-e1026. doi:10.1212/WNL.0000000000008902
41. Yue JK, Yuh EL, Korley FK, et al. Association between plasma GFAP concentrations and MRI abnormalities in patients with CT-negative traumatic brain injury in the TRACK-TBI cohort: a prospective multicentre study. *Lancet Neurol*. 2019;18(10):953-961. doi:10.1016/S1474-4422(19)30282-0

**Table 1: Demographics and Injury Characteristics**

| Variable | TBI (n = 35) Subacute Period | TBI (n = 35) Chronic period | Controls (n = 35) | Statistics |
|---|---|---|---|---|
| Age (μ ± SD) | 34.6 ± 13.8 | | 30.6 ± 9.1 | p = 0.38 (TBI vs. Control) |
| Sex (male/female) | 27/8 | | 21/14 | p = 0.19 (TBI vs. Control) |
| Abnormal head CT (%) | 21 (60%) | | N/A | N/A |
| Extra-axial bleed (%) | 19 (54.2%) | | N/A | N/A |
| Contusion (%) | 12 (34.2%) | | N/A | N/A |
| Skull fracture (%) | 12 (34.2%) | | N/A | N/A |
| GFAP (pg/ml, μ ± SD) | 229.7 (386.4) | 95.6 (68.6) | N/A | p = 0.02 (Subacute vs. Chronic) |
| NfL (pg/ml, μ ± SD) | 132.9 (158.7) | 13.5 (8.8) | N/A | p < 0.001 (Subacute vs. Chronic) |
| BSI score (μ ± SD) | 9.2 (8.0) | 4.8 (5.3) | 1.8 (2.4) | p = 0.01 (Subacute vs. Chronic) |
| RPQ score (μ ± SD) | 19.1 (11.5) | 10.8 (9.2) | 2.5 (4.8) | p = 0.005 (Subacute vs. Chronic) |
| GOSE score (μ ± SD) | 6.3 (1.2) | 7.1 (1) | 8 (0.2) | p = 0.001 (Subacute vs. Chronic) |

*Abbreviations:* BSI, Brief Symptom Inventory; GOSE, Glasgow Outcome Scale Extended; GFAP, glial fibrillary acid protein; N/A, not applicable; NFL, neurofilament light; SD, standard deviation; RPQ, Rivermead Post-Concussive Questionnaire; TBI, traumatic brain injury; μ, mean; SD, standard deviation.

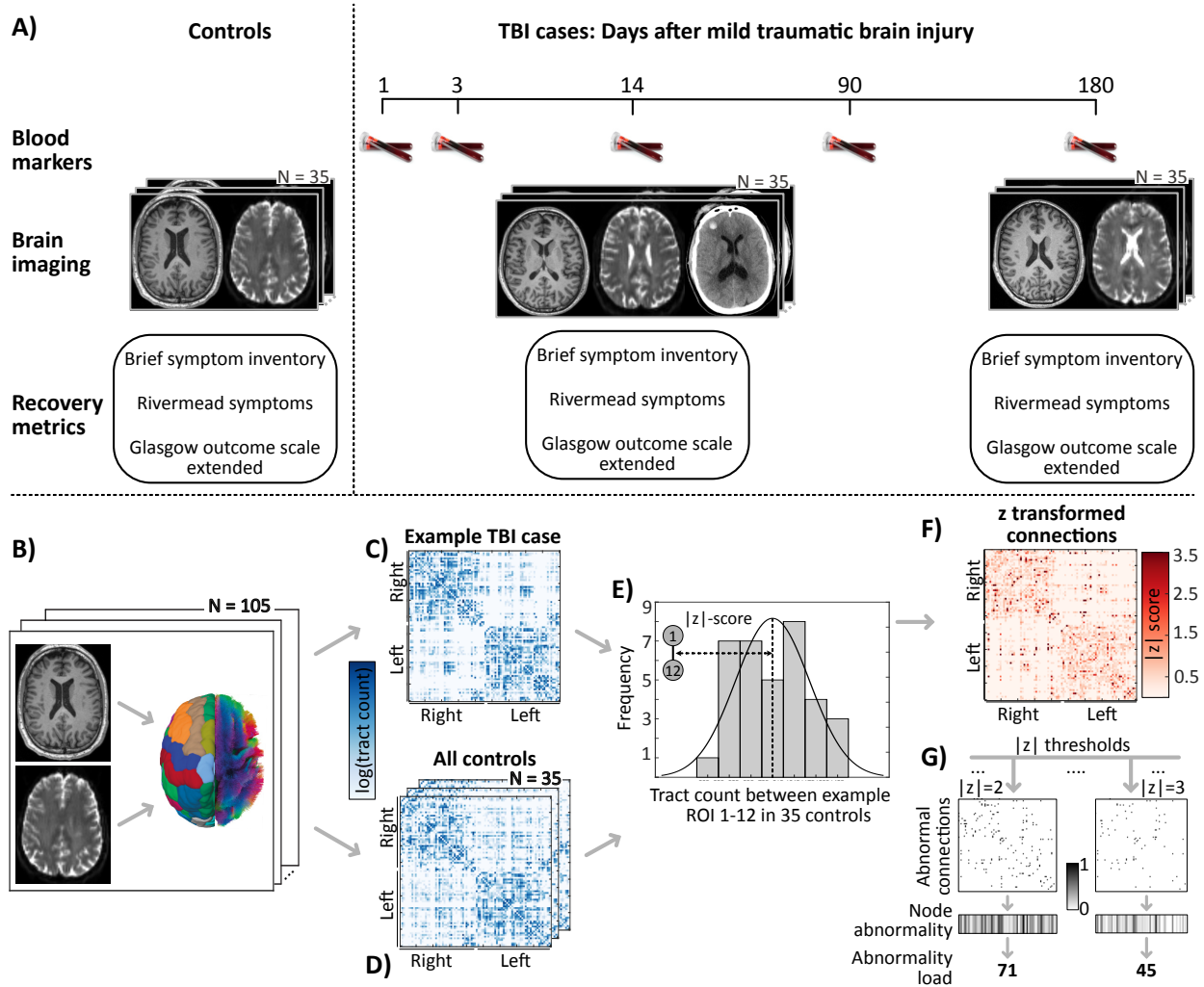

**Figure 1: Overall approach. (A)** Brain imaging, blood draws, and outcome assessment data for post-TBI recovery were collected for 35 controls and 35 TBI cases at different time points after injury. The subacute period was approximately 14 days and the chronic period was approximately 180 days (6 months) after injury. Brain imaging included the T1-weighted and diffusion-weighted MRI for all controls and individuals with TBI during the subacute and chronic periods. CT images were acquired for all TBI cases between hospital admission and the subacute period. Blood plasma concentrations of GFAP and NfL (blood biomarkers of TBI) were estimated at different time points after injury, predominantly during the subacute and chronic periods. **(B)** For each brain imaging scan across all individuals, we constructed a structural brain network consisting of gray matter region/node and the white matter streamlines connecting each region/node. **(C, D)** We generated a connectivity matrix for each scan. The x- and y-axes of the matrix are the nodes

identified by the brain parcellation scheme, and the elements of the matrix are the connections representing the number of white matter streamlines connecting each pair of nodes. **(E, F)** We computed the z-score for every connection by normalizing each connection with the mean and standard deviation of the corresponding connections in uninjured controls. The normalization transformed the connectivity matrix into a z-score abnormality matrix, where a high value indicated a more abnormal connection. **(G)** We assessed abnormal connections at multiple choices of z-score thresholds. We binarized the z-transformed connections at each threshold, with one indicating abnormal connections, when z-score is above the threshold. Node abnormality of each node is the ratio of abnormal connections to that node/region over the total number of connections to that node. Abnormality load is the total number of abnormal nodes at any choice of z-score threshold.

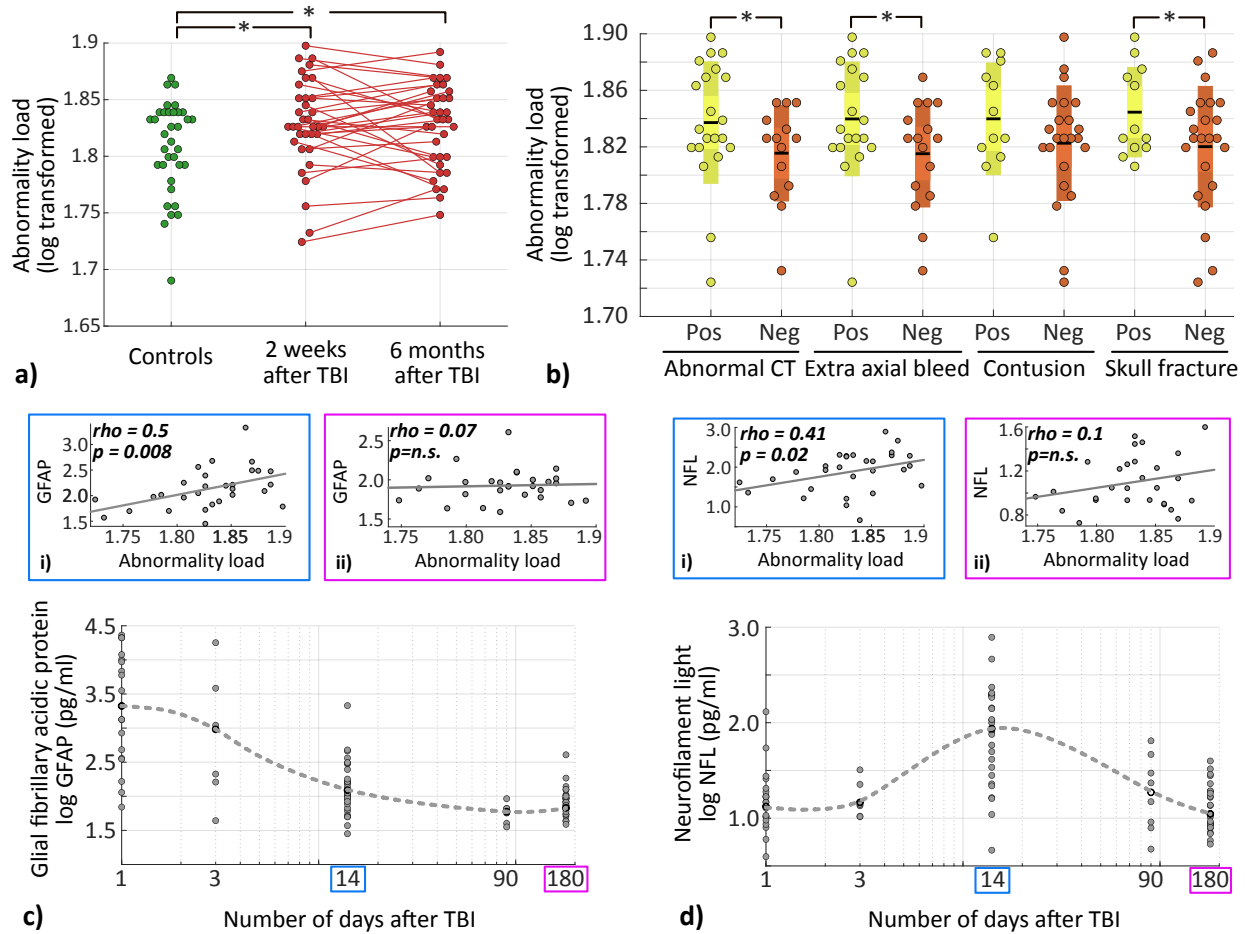

**Figure 2: Abnormality load, acute CT findings, and blood biomarkers. a)** Individuals with TBI have significantly higher abnormality compared to controls both in the subacute (2 weeks: p = 0.03, d = 0.51) and chronic (6 months: p = 0.04, d = 0.48) periods after the injury. In individual cases, the abnormality load can either increase, decrease, or remain the same between the subacute and chronic periods, indicated by the connecting lines. **b)** Presence of acute intracranial lesions on CT was associated with significantly higher abnormality load than individuals who were CT negative after brain injury, suggesting that abnormality may increase with injury severity (abnormal CT: p = 0.05, d = 0.55; extra axial bleed: p = 0.03, d = 0.63; contusion: p = 0.11, d = 0.43; skull fracture: p = 0.03, d = 0.62). Panel **c)** plots the trajectory of the expression of GFAP at different time points of the blood draw, and panel **d)** plots the equivalent trajectory for the expression of NfL. These blood biomarkers of injury are expressed highly at different time points after the injury and gradually subside. The inset plots **(i)** in the blue box shows the correlation between the expression of the glial fibrillary acid protein (GFAP) and neurofilament light (NFL) in the subacute period and inset **(ii)** in the pink box shows the correlation between abnormality

load and the expression of GFAP and NFL in the chronic period. Increase in abnormality load during the subacute period is correlated with the increase in expression of both blood protein biomarkers in the subacute period (GFAP vs. abnormality load: r = 0.50, p = 0.008; NFL vs. abnormality load: r = 0.41, p = 0.02).

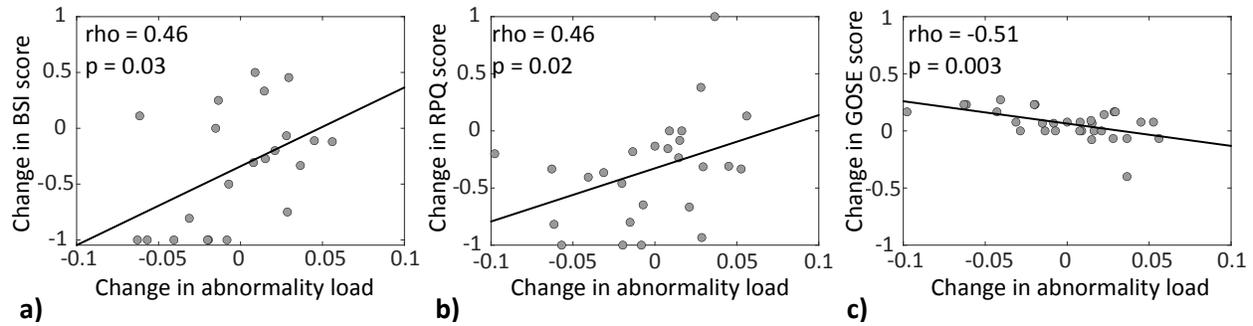

**Figure 3: Abnormality load correlations with outcomes.** The proportional change in abnormality load from the subacute to chronic post-injury period correlated with the proportional change in measures of **a)** subjective psychological symptoms (BSI: r = 0.46, p = 0.03), **b)** post-concussive symptoms (RPQ: r = 0.46, p = 0.02), and **c)** functional outcome (GOSE: r = -0.51, p = 0.003). Recovery is indicated by a decrease in BSI and RPQ scores and increase in GOSE score. On proportional change scale, negative values indicate decreases while positive values indicate increase. The correlations indicate that increases in structural network abnormality over time correlates with worsening psychological and post-concussive symptoms as well as functional outcome over time. BSI, Brief Symptom Inventory; RPQ, Rivermead Post-Concussive Questionnaire; GOSE, Glasgow Outcome Scale Extended.

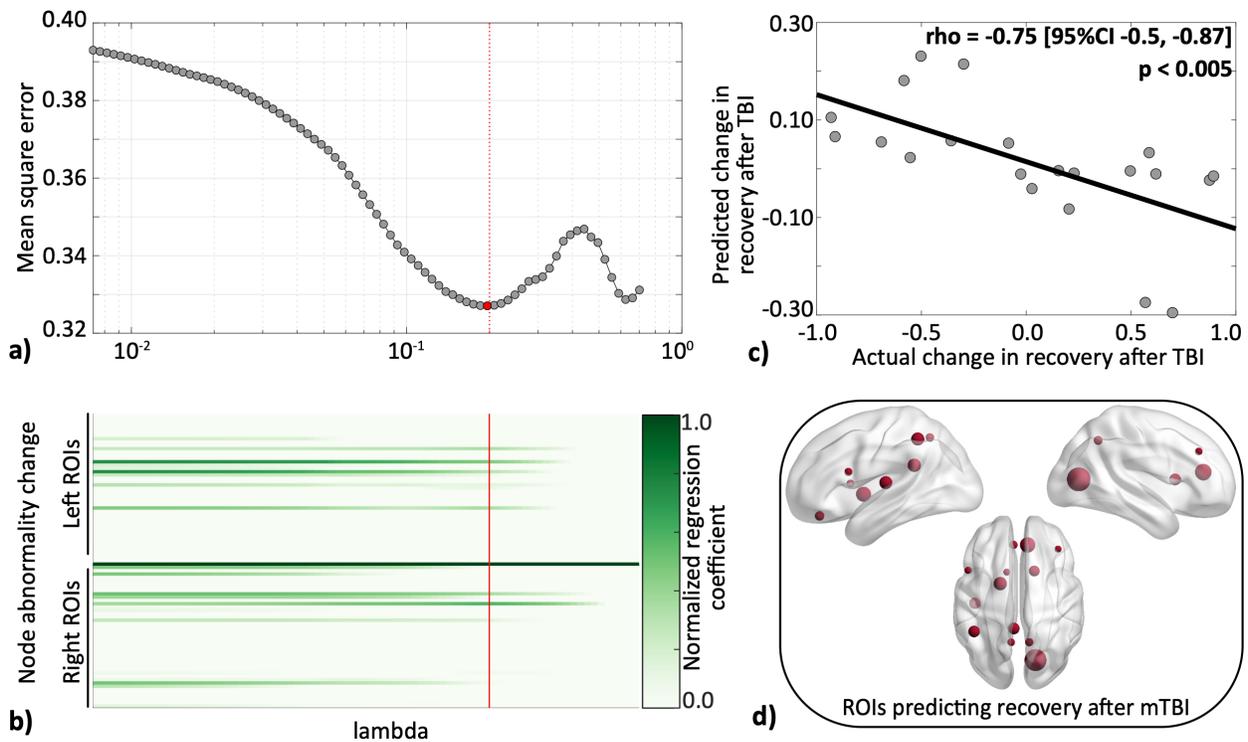

**Figure 4: Node abnormality and outcome.** Panel **a)** shows the mean square error for different values of regularization parameter lambda in the elastic net regression model. Panel **b)** shows the relative feature importance of brain areas corresponding to varying lambda. The red vertical line indicates the lambda value at which the lowest mean square error was achieved in an optimal elastic net regression model. This optimal regression model assigned non-zero coefficients to 14 brain areas as plotted in horizontal green lines. Panel **c)** shows the prediction performance of the optimal model in predicting change in recovery status. The predicted performance from the regression model is plotted against the actual change in recovery. The high correlation coefficient between predicted and actual recovery changes (r = -0.75 [ 95% CI -0.5 -0.87], p < 0.005) suggests that monitoring the change in abnormality for the brain regions in panel **d)** predicts patient recovery after TBI. Panel **d)** plots the spatial location of these 14 brain regions at which the proportional change in abnormality load were selected as the salient features to predict proportional change in recovery status. Sphere sizes are drawn to scale with the relative feature importance.

# Supplementary: Change in structural brain network abnormalities after traumatic brain injury determines post-injury recovery

James J Gugger*,[1,2], Nishant Sinha*,[1,2], Yiming Huang[3], Alexa Walter[1], Cillian Lynch[1], Justin Morrison[1], Nathan Smyk[1], Danielle Sandsmark[1], Ramon Diaz-Arrastia†[1], Kathryn A Davis†[1,2]

*Contributed equally as the first author.
†Contributed equally as senior author.

[1]Department of Neurology, Perelman School of Medicine, University of Pennsylvania, Philadelphia, PA, USA

[2]Center for Neuroengineering & Therapeutics, University of Pennsylvania, Philadelphia, PA, USA

[3]Interdisciplinary Computing and Complex BioSystems, School of Computing, Newcastle University, Newcastle upon Tyne, United Kingdom

Correspondence to: Nishant Sinha
Address: Hayden Hall Room 306, University of Pennsylvania, 240 South 33rd Street, Philadelphia, PA 19104, USA
Email: nishant.sinha89@gmail.com
Twitter: @_Nishant_Sinha
Orcid ID: 0000-0002-2090-4889

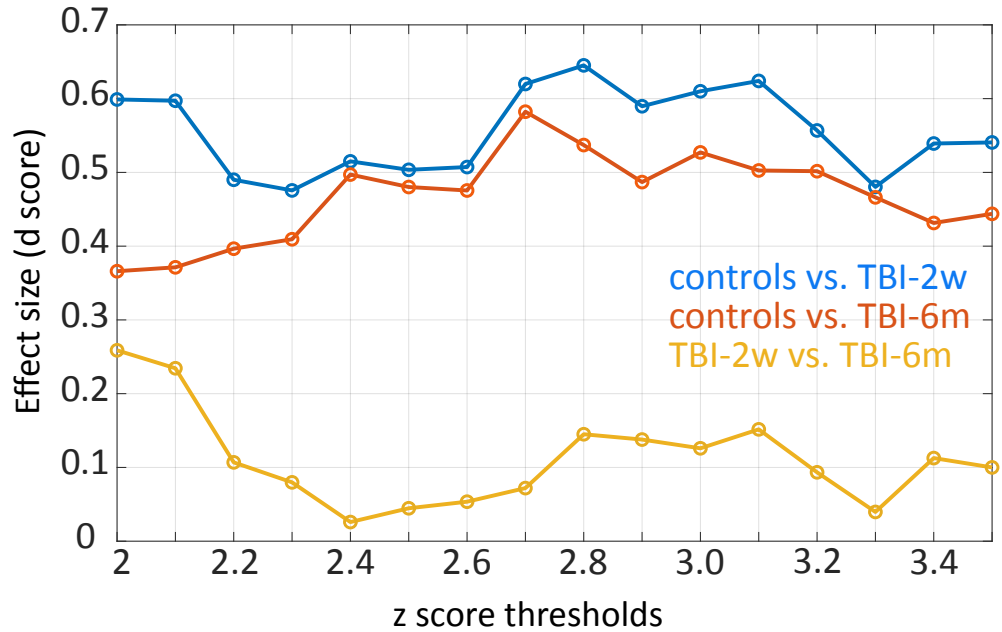

**Figure S1: Increased abnormality load in individuals with TBI compared to controls is consistent across the choice of z-score thresholds.** Equivalent to Figure 2 in the main manuscript, where we illustrated our findings for one example threshold at z > 2.5, this plot shows consistent findings for a range of z-score thresholds. The plot in blue shows the effect size of the abnormality load between controls and individuals with TBI in the subacute period (2 weeks), and the plot in red shows the same for the chronic period (6 months) across different choices of z-score thresholds. At every z-score threshold choice, we found *p < 0.05* after FDR correction. The plot in yellow shows paired Cohen's d score for change in abnormality load in TBI cases between subacute and chronic periods. Consistent with our results in the main manuscript, we did not detect a significant change in abnormality load in TBI cases from subacute to chronic periods at a group level.

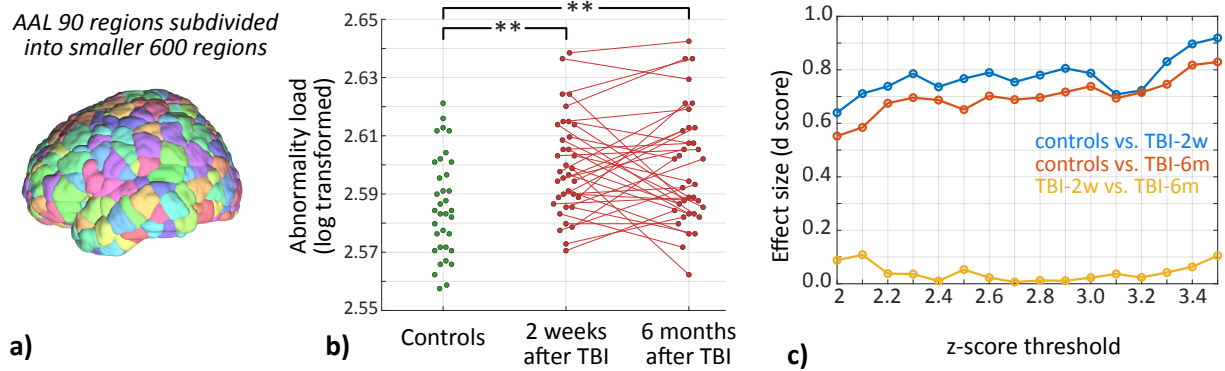

**Figure S2: Increased abnormality load in individuals with TBI compared to controls is consistent with a different choice of brain parcellation scheme.** Panel **a)** shows AAL atlas with 600 regions of approximately equal size. These 600 regions are subdivisions of the original AAL 90 regions. We repeated our analysis using this new atlas to verify consistency across the choice of parcellation scheme. Equivalent to our results in the main manuscript, panel **b)** shows that individuals with TBI have higher abnormality compared to controls both in the subacute (2 weeks: d = 0.78, p = 0.001) and chronic (6 months: d = 0.70, p = 0.003) periods after TBI. z-score threshold was 2.6. Equivalent to Figure S1, panel **c)** shows consistent results across the choice of z-score thresholds in this alternate brain parcellation scheme.

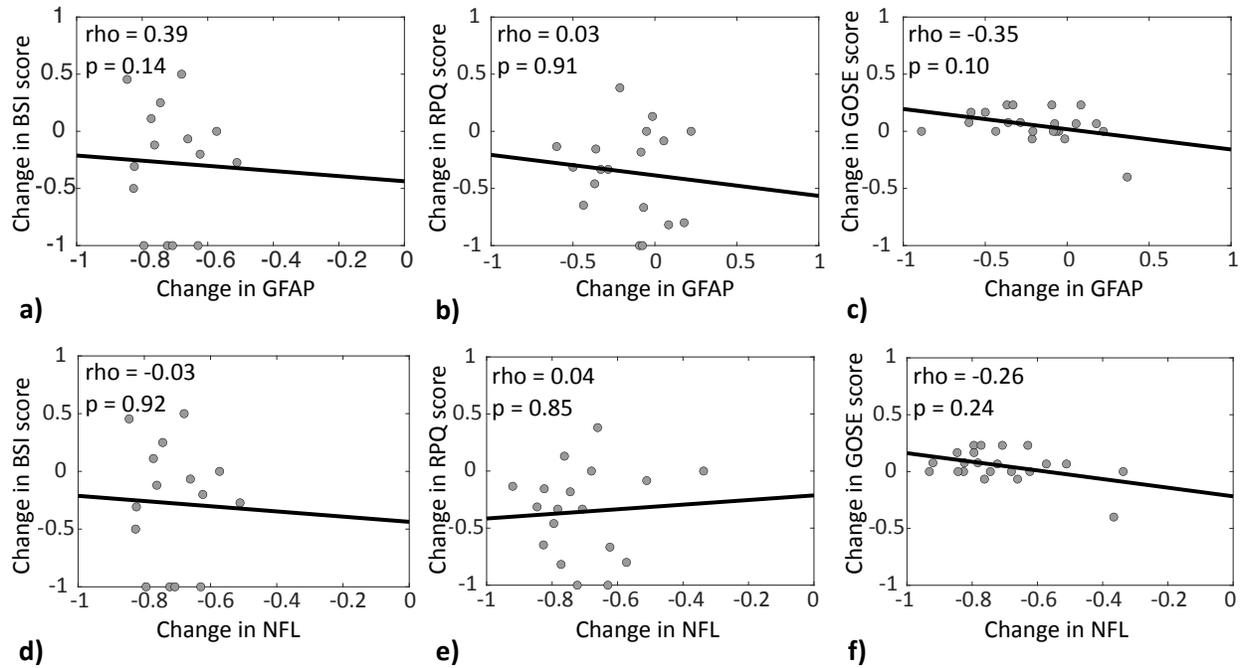

**Figure S3: No association between proportional change in RPQ, BSI, and GOSE score with a proportional change in GFAP and NFL.** The change in blood plasma concentration of GFAP (panels a-c) and NFL (panels d-f) from the subacute to chronic periods did not correlate with the change in subjective psychological symptoms (BSI score), post-concussive symptoms (RPQ), or functional outcome (GOSE score) from the subacute to chronic post-injury period.

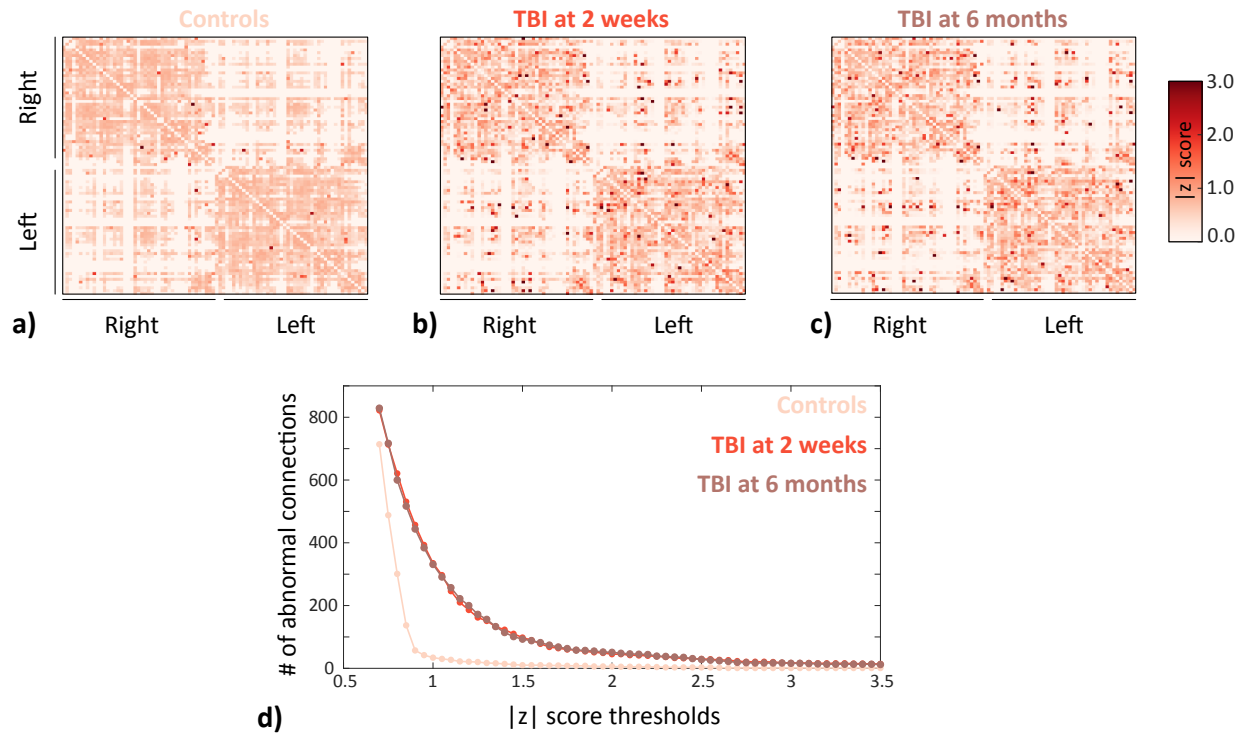

**Figure S4: Average z-transform connectivity matrix for controls and individuals with TBI at subacute (2 weeks) and chronic (6 months) periods post-injury. a-c)** Individuals with TBI exhibit many abnormal connections compared to controls. More abnormal connections in darker red color are apparent for TBI individuals compared to the control group. **d)** At each choice of z-score threshold, we binarized the connectivity matrices for the three groups and counted the number of abnormal connections. The plot shows that for each choice of the z-score threshold, the number of abnormal connections is more in individuals with TBI than in controls.

**Supplementary Table S1:**

Structural network where proportional change in abnormality load predicts proportional change in recovery status.

| Normalized coefficient* | Region of Interest |
|---|---|
| 1.00 | Right Calcarine |
| 0.58 | Right Anterior Cingulate |
| 0.45 | Left Pallidum |
| 0.35 | Left Posterior Cingulate |
| 0.34 | Left Inferior Parietal |
| 0.33 | Left Heschl |
| 0.29 | Right Caudate |
| 0.14 | Left Gyrus Rectus |
| 0.13 | Right Precuneus |
| 0.09 | Left Precuneus |
| 0.06 | Left Inferior Frontal Operculum |
| 0.04 | Left Caudate |
| 0.04 | Right Middle Frontal |
| 0.01 | Right Mid Cingulate |

*Coefficients are normalized so that the highest feature importance is equal to 1 and the rest of the coefficients are scaled from 0-1.